\documentclass[twocolumn]{aastex61}
\usepackage{amsfonts,amsmath,amstext,graphicx,apjfonts,subfigure,color,breakcites}

\usepackage{apjfonts} 

\def\ra#1#2#3{#1$^{\rm h}$ #2$^{\rm m}$ #3$^{\rm s}$}
\def\dec#1#2#3{$#1^\circ #2' #3''$}

\def\swift{{\it Swift}}
\def\chandra{{\it Chandra}}
\def\xmm{{\it XMM-Newton}}
\def\gw{GW170817}
\def\ngc{NGC\,4993}

\shorttitle{GW\,170817}
\shortauthors{Fong et al.}

\begin{document}

\title{The Electromagnetic Counterpart of the Binary Neutron Star Merger LIGO/VIRGO GW170817: \\ VIII. A Comparison to Cosmological Short-duration Gamma-ray Bursts}

\author{W.~Fong}
\altaffiliation{Hubble Fellow}
\affiliation{Center for Interdisciplinary Exploration and Research in Astrophysics (CIERA) and Department of Physics and Astronomy, Northwestern University, Evanston, IL 60208}

\author{E.~Berger}
\affiliation{Harvard-Smithsonian Center for Astrophysics, 60 Garden Street, Cambridge, Massachusetts 02138, USA}

\author{P.~K.~Blanchard}
\affiliation{Harvard-Smithsonian Center for Astrophysics, 60 Garden Street, Cambridge, Massachusetts 02138, USA}

\author{R.~Margutti}
\affiliation{Center for Interdisciplinary Exploration and Research in Astrophysics (CIERA) and Department of Physics and Astronomy, Northwestern University, Evanston, IL 60208}

\author{P.~S.~Cowperthwaite}
\affiliation{Harvard-Smithsonian Center for Astrophysics, 60 Garden Street, Cambridge, Massachusetts 02138, USA}

\author{R.~Chornock}
\affiliation{Astrophysical Institute, Department of Physics and Astronomy, 251B Clippinger Lab, Ohio University, Athens, OH 45701, USA}

\author{K.~D.~Alexander}
\affiliation{Harvard-Smithsonian Center for Astrophysics, 60 Garden Street, Cambridge, Massachusetts 02138, USA}

\author{B.~D.~Metzger}
\affiliation{Department of Physics and Columbia Astrophysics Laboratory, Columbia University, New York, NY 10027, USA}

\author{V.~A.~Villar}
\affiliation{Harvard-Smithsonian Center for Astrophysics, 60 Garden Street, Cambridge, Massachusetts 02138, USA}

\author{M.~Nicholl}
\affiliation{Harvard-Smithsonian Center for Astrophysics, 60 Garden Street, Cambridge, Massachusetts 02138, USA}

\author{T.~Eftekhari}
\affiliation{Harvard-Smithsonian Center for Astrophysics, 60 Garden Street, Cambridge, Massachusetts 02138, USA}

\author{P.~K.~G.~Williams}
\affiliation{Harvard-Smithsonian Center for Astrophysics, 60 Garden Street, Cambridge, Massachusetts 02138, USA}

\author{J.~Annis}
\affiliation{Fermi National Accelerator Laboratory, P.O. Box 500, Batavia, IL 60510, USA}

\author{D.~Brout}
\affiliation{Department of Physics and Astronomy, University of Pennsylvania, Philadelphia, PA 19104, USA}

\author{D.~A.~Brown}
\affiliation{Department of Physics, Syracuse University, Syracuse NY 13224, USA}

\author{H.-Y.~Chen}
\affiliation{Kavli Institute for Cosmological Physics, University of Chicago, Chicago, IL 60637, USA}

\author{Z.~Doctor}
\affiliation{Kavli Institute for Cosmological Physics, University of Chicago, Chicago, IL 60637, USA}

\author{H.~T.~Diehl}
\affiliation{Fermi National Accelerator Laboratory, P.O. Box 500, Batavia, IL 60510, USA}

\author{D.~E.~Holz}
\affiliation{Enrico Fermi Institute, Department of Physics, Department of Astronomy and Astrophysics}
\affiliation{Kavli Institute for Cosmological Physics, University of Chicago, Chicago, IL 60637, USA}

\author{A.~Rest}
\affiliation{Space Telescope Science Institute, 3700 San Martin Drive, Baltimore, MD 21218, USA}
\affiliation{Department of Physics and Astronomy, The Johns Hopkins University, 3400 North
Charles Street, Baltimore, MD 21218, USA}

\author{M.~Sako}
\affiliation{Department of Physics and Astronomy, University of Pennsylvania, Philadelphia, PA 19104, USA}

\author{M.~Soares-Santos}
\affiliation{Fermi National Accelerator Laboratory, P.O. Box 500, Batavia, IL 60510, USA}
\affiliation{Department of Physics, Brandeis University, Waltham, MA 02452, USA}

\begin{abstract}
We present a comprehensive comparison of the properties of the radio through X-ray counterpart of \gw\ and the properties of short-duration gamma-ray bursts (GRBs). For this effort, we utilize a sample of 36~short GRBs spanning a redshift range of $z \approx 0.12-2.6$ discovered over 2004-2017. We find that the counterpart to \gw\ has an isotropic-equivalent luminosity that is $\approx 3000$~times less than the median value of on-axis short GRB X-ray afterglows, and $\gtrsim10^{4}$ times less than that for detected short GRB radio afterglows. Moreover, the allowed jet energies and particle densities inferred from the radio and X-ray counterparts to \gw\ and on-axis short GRB afterglows are remarkably similar, suggesting that viewing angle effects are the dominant, and perhaps only, difference in their observed radio and X-ray behavior. From comparison to previous claimed kilonovae following short GRBs, we find that the optical and near-IR counterpart to \gw\ is comparatively under-luminous by a factor of $\approx 3-5$, indicating a range of kilonova luminosities and timescales. A comparison of the optical limits following short GRBs on $\lesssim 1$~day timescales also rules out a ``blue'' kilonova of comparable optical isotropic-equivalent luminosity in one previous short GRB. Finally, we investigate the host galaxy of \gw, \ngc, in the context of short GRB host galaxy stellar population properties. We find that \ngc\ is superlative in terms of its large luminosity, old stellar population age, and low star formation rate compared to previous short GRB hosts. Additional events within the Advanced LIGO/Virgo volume will be crucial in delineating the properties of the host galaxies of NS-NS mergers, and connecting them to their cosmological counterparts.
\end{abstract}

\section{Introduction}

Short-duration gamma-ray bursts (GRBs) have long been linked to compact object merger progenitors \citep{elp+89,npp92}, involving two neutron stars (NS-NS) or a neutron star and a black hole (NS-BH). This link has been strengthened by a wealth of indirect observational evidence: the lack of associated supernovae to deep optical limits \citep{ffp+05,hsg+05,hwf+05,sbk+06,ktr+10,rwl+10,ber14,fbm+14}, the observed optical/near-IR excess emission following some short GRBs interpreted as $r$-process kilonovae \citep{pmg+09,bfc13,tlf+13,jlc+15,yjl+15,jhl+16}, the locations of short GRBs within their host galaxies which are well-matched to predictions for NS-NS mergers  \citep{fbf10,fb13}, and the sizable fraction of short GRBs that occur in early-type galaxies \citep{bpc+05,gso+05,bpp+06,fbc+13}, indicative of older stellar progenitors \citep{zr07}. While this indirect evidence has been strongly in favor of NS-NS/NS-BH merger progenitors, to date there has been no direct evidence linking short GRBs to their origin.

The Advanced Laser Interferometer Gravitational-Wave Observatory (LIGO) and Advanced Virgo announced the detection of a gravitational wave candidate on 2017 August 17 at 12:41:04 UT with a high probability of being a NS-NS merger \citep{ALVgcn,ALVdetection}. At 12:41:06.47 UT, a weak, gamma-ray transient with a duration of $\sim 2$~sec was discovered by the Gamma-ray Burst Monitor (GBM) on-board the {\it Fermi} satellite \citep{GBMdetection}, with a delay of $\approx 2$~s from the aLIGO/Virgo trigger time. The near-coincident detection of a gamma-ray transient and a neutron star merger detected by gravitational waves potentially provides the first ``smoking gun'' evidence that at least some short GRBs originate from neutron star mergers.

Following the detection, our search with the Dark Energy Camera (DECam) yielded the detection of an optical transient located at RA=\ra{13}{09}{48.08} and Dec=\dec{-23}{22}{53.2} (J2000; \citealt{Marcelle}), situated $\sim10''$ from \ngc\ \citep{hms+91}. This fading optical transient\footnote{This source is also known as AT2017gfo (International Astronomical Union name), SSS17a \citep{SWOPEgcn,SWOPEpaper}, and DLT17ck \citep{DLT40gcn,DLT40Paper}.} was demonstrated to be the likely optical counterpart to \gw\ from its temporal evolution \citep{Phil}. We initiated observations across the electromagnetic spectrum, and discovered a brightening X-ray source \citep{Raff}, monitored the evolution of the transient with optical \citep{Matt} and near-IR spectroscopy \citep{Chornock}, and detected the radio counterpart at GHz frequencies \citep{Kate}. The details and interpretation of each data set are described in the individual papers.

Here, we leverage our extensive data set of the electromagnetic counterpart to \gw\ to provide a broader picture of the event in the context of the cosmological short GRB population. In particular, we examine the counterpart with respect to short GRB afterglows, searches for kilonovae following short GRBs, and short GRB locations with respect to their host galaxies. We also examine the host galaxy \ngc\ with respect to the stellar population properties of galaxies hosting short GRBs.

Unless otherwise noted, all magnitudes in this paper are in the AB system and reported uncertainties correspond to $68\%$ confidence. We employ a standard $\Lambda$CDM cosmology with $\Omega_M=0.27$, $\Omega_\Lambda=0.73$, and $H_0=71.0$ km s$^{-1}$ Mpc$^{-1}$.  We note that all luminosities quoted in this paper correspond to isotropic-equivalent values. We adopt a distance to both \ngc\ and the counterpart of $D_L=39.5$~Mpc \citep{hms+91}.

\section{Sample \& Observations}

We collect observations of previous, well-localized short GRBs across the electromagnetic spectrum, as well as observations of their host galaxies.

\subsection{Redshifts}

\begin{figure}
\includegraphics*[width=0.48\textwidth,clip=]{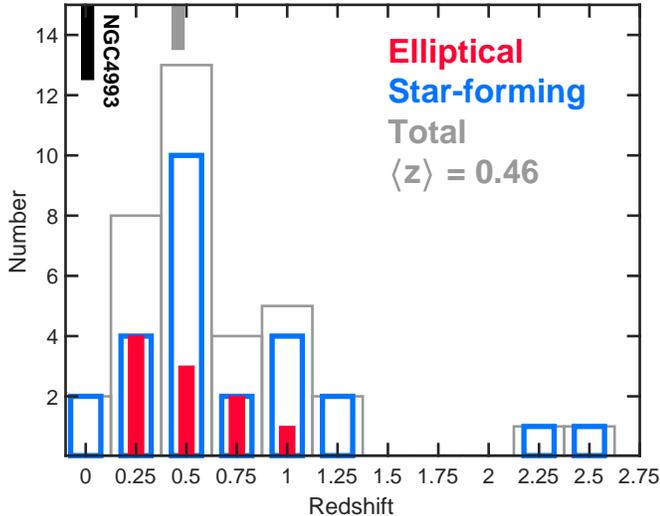} 
\caption{Redshift distribution of 36 short GRBs discovered over 2004-2017 (grey). The distributions are divided by host galaxy type: elliptical (red) and star-forming (blue) galaxies. The redshift of NGC4993 is denoted (black line), along with the median for the short GRB population (grey line) of $\langle z \rangle \approx 0.46$.
\label{fig:zhist}}
\end{figure}

To enable an effective comparison to the counterpart to \gw, we utilize data from all previous short GRBs with redshift measurements. We compile redshift measurements from \citet{ber14} for bursts discovered over 2004-2013. To update this sample, we also include redshifts for new bursts from GCN circulars \citep{hms+14,cl16,lwt+16}, recent papers on individual events \citep{fmc+16,tsc+16,skm+17}, and additional optical spectroscopy which will be described in an upcoming work (Fong~et~al. in prep). Our total sample of short GRBs with redshifts comprises 36 events and spans a redshift range of $\approx 0.12-2.6$, with a median of $\langle z \rangle \approx 0.46$ (Figure~\ref{fig:zhist}). 

We note that, by default, requiring that the events have redshifts only includes bursts that are well-localized to $\lesssim$few arcsec uncertainty via the detection of an X-ray afterglow. However, we previously explored any potential bias within the \swift\ short GRB population introduced by requiring an X-ray afterglow based on a smaller sample of events \citep{fbc+13}. We found that the large majority of bursts that did not have detected X-ray afterglows were affected by observing constraints that would generally be decoupled from any burst or host galaxy properties. Thus, we do not expect the requirement of a redshift to introduce any discernible bias in this sample.

%%%%%%%%%%%%%%% FIGURE
%%%%%%%%%%%%%%% FIGURE

\subsection{X-rays}

In order to compare the X-ray emission of on-axis short GRB afterglows to the X-ray counterpart for \gw\ \citep{SWIFTpaper,Raff}, we utilize the sample in \citet{fbm+15} which covers short GRBs discovered between November 2004 and March 2015 (references for individual data sets therein). The data were taken primarily with the X-ray Telescope (XRT; \citealt{bhn+05}) on-board the {\it Swift} satellite \citep{ggg+04}, the {\it Chandra X-ray Observatory} and \xmm. To supplement this data set, we gather all \swift/XRT observations from the light curve repository \citep{ebp+07,ebp+09} for short GRBs with detected afterglows between March 2015 and August 2017. To enable a direct comparison to the X-ray counterpart to \gw, we use the burst redshifts to convert to X-ray luminosity. Our sample of short GRBs with X-ray afterglows comprises 36~events, and the light curves are shown in Figure~\ref{fig:ag}.

\subsection{Radio}

\begin{figure*}
\begin{minipage}[c]{\textwidth}
\tabcolsep0.0in
\includegraphics*[width=0.48\textwidth,clip=]{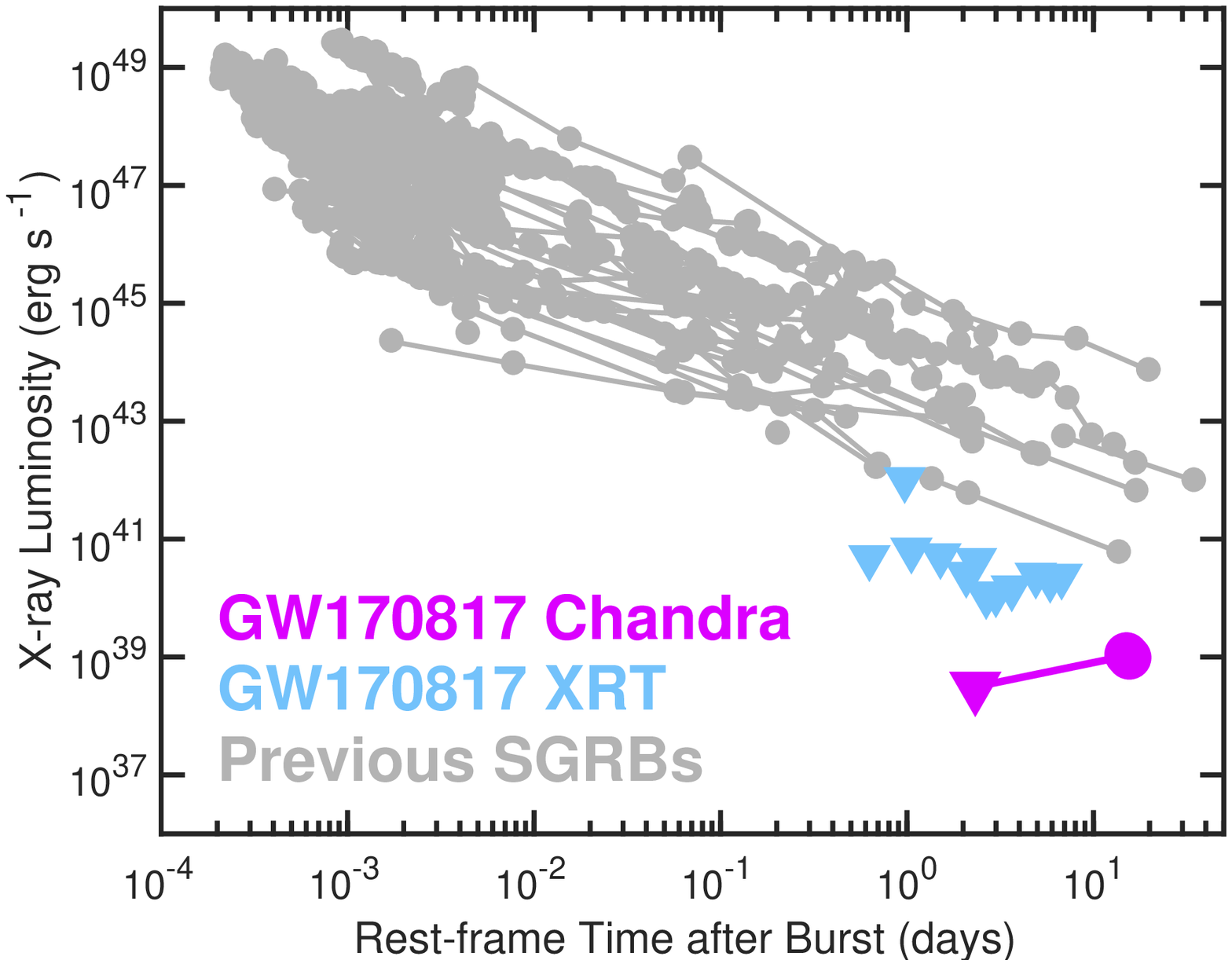} 
\includegraphics*[width=0.48\textwidth,clip=]{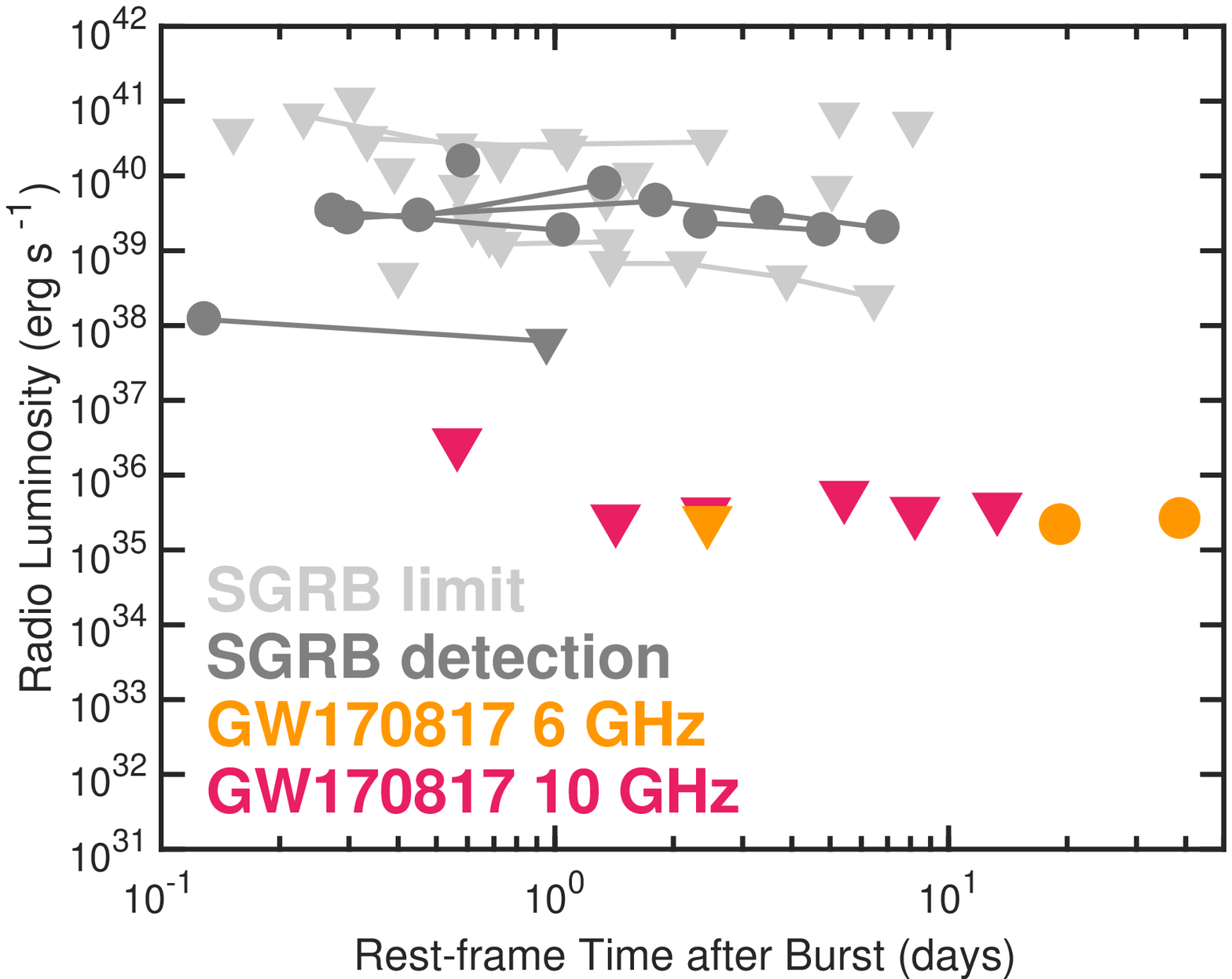}
\end{minipage}
\caption{{\it Left:} Light curve of the X-ray counterpart to \gw\ from \swift\ and \chandra\ ($0.3-10$~keV; \citealt{DarylCircular,Daryl,Raff}, where circles denote detections and triangles denote $3\sigma$ upper limits). Also shown are on-axis X-ray afterglow light curves from all previous short GRBs with well-sampled X-ray light curves and redshifts, comprising 36 events. At the time of detection, \gw\ has an isotropic-equivalent luminosity that is $\approx 3000$ times less than the median short GRB X-ray afterglow, and $\approx 50$ times less than the faintest detected X-ray emission from a short GRB. {\it Right:} Radio search for the counterpart of \gw\ from \citet{Kate} at 6~GHz (orange) and 10~GHz (red) with the VLA, yielding detections at 6~GHz beyond $\approx 19$~days. The six short GRBs with radio afterglow detections (dark grey circles) along with $3\sigma$ upper limits for 19 additional events with redshifts (light grey triangles) are shown. These observations demonstrate that the radio counterpart to \gw\ is $\gtrsim 10^{4}$ times less luminous (isotropic-equivalent) than detected radio afterglows at similar epochs, and $\gtrsim 500$ times less luminous than the faintest detected radio afterglow for a short GRB.
\label{fig:ag}}
\end{figure*}

We compile radio observations for short GRBs in a similar manner to that for the X-ray observations, starting with the sample from \citet{fbm+15}. We update this sample to include five additional events with observations from the Karl G. Jansky Very Large Array (VLA) under Programs 15A-235 (PI:~Berger) and 17A-218 (PI:~Fong). Of the bursts with redshifts, four have previously-reported detections \citep{bpc+05,sbk+06,fbm+14,fbm+15} while two additional detections are included in the updated sample. In total, the sample includes six bursts with detections and 19 additional events with upper limits. We convert the data to luminosity via the burst redshifts. The resulting radio light curves and upper limits, along with the results of radio monitoring following \gw\ at $6$~GHz and $10$~GHz with the VLA \citep{Kate}, are shown in Figure~\ref{fig:ag}.

%%%%%%%%%%%%%%% FIGURE

%%%%%%%%%%%%%%% FIGURE

\subsection{Optical/Near-IR Kilonovae}

There have been three previous claims of kilonovae in the literature: a near-IR excess following GRB\,130603B ($z=0.356$; \citealt{bfc13,tlf+13}), and optical excesses following GRB\,050709 ($z=0.16$; \citealt{jhl+16}) and GRB\,060614\footnote{GRB\,060614 had a duration of $\approx 108$~seconds and would typically be classified as a long-duration GRB. However, this event lacks an associated supernova to deep limits, suggesting that it does not have a massive star progenitor \citep{gfp+06}, and thus we include it in this discussion.} ($z=0.125$; \citealt{jlc+15,yjl+15}). We note an additional optical excess was reported following GRB\,080503 \citep{pmg+09}; however, since this burst does not have a known redshift, we do not include it in this discussion. To enable a comparison of the luminosity and temporal behavior of the optical/near-IR counterpart to \gw\ \citep{Phil,Chornock,Matt}, we collect observations corresponding as close as possible to the rest-frame $r$- and $H$-bands for each burst \citep{bfc13,tlf+13,jlc+15,jhl+16}. 

We supplement these detections with any optical/near-IR observations following short GRBs on timescales of $\gtrsim\!1$~day after the burst. To this end, we retrieve {\it Hubble Space Telescope} ({\it HST}) observations (PI:~Tanvir; Program 14237) of the short GRB\,160821B from the Mikulski Archive for Space Telescopes (MAST) archive. We utilize observations taken with the Wide Field Camera~3  (WFC3) in the F160W filter, corresponding to rest-frame $H$-band at the redshift of the burst ($z=0.16$; \citealt{lwt+16}), on 2016 September 14 UT ($\sim 23$~days after the burst). We used the {\tt astrodrizzle} task as part of the Drizzlepac software package \citep{gon12} to create final drizzled image, using {\tt final\_scale} = $0.065''$ pixel$^{-1}$ and {\tt final\_pixfrac} = 0.8. We use standard tasks in IRAF to perform aperture photometry of faint point sources in the field to calculate a $3\sigma$ limit of $m_{\rm 160W} \gtrsim$26.0~mag.

We add to this sample optical and near-IR upper limits following 14 additional events with redshifts. Compared to the sample in \citet{fbm+15}, we note the addition of a deep limit following GRB\,050509B of $r \gtrsim 25.7$~mag at $\approx 25.9$~hr after the burst \citep{csb+05,bpp+06}. Although this limit corresponds to rest-frame $V$-band, we note that the expected $V-r$ color based on observations of the optical counterpart to \gw\ at $\approx 1.5$~days after the event \citep{Phil} is negligible, $\lesssim 0.2$~mag. Thus, we still include this point in our sample.

For all bursts with detections of or limits on kilonova emission, we use the burst redshifts to convert apparent magnitude to $K$-corrected absolute magnitudes. The data for the previous short GRBs, as well as Gemini-South $H$-band observations of the counterpart to \gw\ \citep{Phil} are shown in Figure~\ref{fig:kn}. To enable a direct comparison to early short GRB optical limits, we also employ the initial DECam observation at $i$-band at $\approx 0.47$~days, and correct for the observed $r-i$ color of $\approx 0.2$~mag at $\approx 1.4$~days \citep{Phil}. We note that this is conservative as the source has a blue color at $\lesssim 1$~day and $0.2$~mag is likely an upper limit on the $r-i$ color. 

\subsection{Host galaxy properties}

To obtain a complete sample of short GRB host properties, we collect all available values for host redshifts, galaxy type, rest-frame $B$-band luminosity ($L_B$), stellar mass ($M_*$), stellar population age ($\tau$), and star formation rate (SFR) from large samples \citep{lb10,fbc+13,ber14} as well as papers on individual objects \citep{pmm+11,dtr+14,fmc+16,tsc+16,skm+17}. We supplement this sample with values for the redshifts and stellar population properties of eight additional short GRB hosts discovered over 2014-2017, the details of which will be described in an upcoming work (Fong~et~al.~in~prep.). For these bursts, we infer stellar masses from a combination of broad-band SED fitting, and $K$-band luminosity and $B-K$ color relations \citep{bj01,mdp+05}. We calculate rest-frame $B$-band luminosities from host photometry. We normalize these values to the characteristic luminosity ($L^*$) of the evolving galaxy luminosity function over $z \sim 0$ to $z\approx 2$ depending on the redshift of each host \citep{bhb+03,wfk+06,mvq+07}. The median of each stellar population property is listed in Table~\ref{tab:props}. The total sample of 36~events provides a comprehensive comparison sample for \ngc.

\section{Comparison to Short GRBs}

\subsection{Afterglows \& Explosion Properties}

We first compare the radio and X-ray observations of the counterpart to \gw\ to those following on-axis short GRBs. In the X-ray band, the counterpart is observed to brighten over $\approx 2.4-15.4$~days, indicative of a weak on-axis jet or an off-axis afterglow observed at an angle $\theta_{\rm obs}$ from the jet axis \citep{Raff}. At the time of the first detection reported in \citet{Raff} at $\approx\!15.4$~days, the X-ray counterpart has a luminosity of $\approx 1.1 \times 10^{39}$~erg~s$^{-1}$ ($0.3-10$~keV). The median luminosity of the five short GRBs that have X-ray detections at similar rest-frame times is $\approx 3 \times 10^{42}$~erg~s$^{-1}$ (Figure~\ref{fig:ag}); thus, the X-ray counterpart to \gw\ is $\approx 3000$ times less luminous. Moreover, the X-ray counterpart to \gw\ is $\approx 50$ times less luminous than the faintest known X-ray emission from a short GRB. Given that the depths of late-time observations following short GRBs are essentially at the limits of current X-ray observatories, Figure~\ref{fig:ag} demonstrates that the counterpart to \gw, if shifted to the redshifts of short GRBs, would not have been detected.

The radio afterglows of short GRBs provide a complementary comparison sample. \citet{Kate} report monitoring of the radio counterpart to \gw\ spanning $\approx\!0.6-40$~days, with deep limits of $\gtrsim (2-5) \times 10^{35}$~erg~s$^{-1}$ until $\approx\!19$~days and detections of $\approx (2.1-2.6) \times 10^{35}$~erg~s$^{-1}$ thereafter (Figure~\ref{fig:ag}). Overall, the radio counterpart to \gw\ is a factor of $\gtrsim\!10^{4}$ times fainter than detected radio afterglows, which have a median of $\approx 4 \times 10^{39}$~erg~s$^{-1}$. The faintest reported detection of an on-axis radio afterglow was following one of the most nearby known short GRBs, GRB\,160821B ($z=0.16$), with $\approx 1.2 \times 10^{38}$~erg~s$^{-1}$ (Figure~\ref{fig:ag}), $\approx500$ times more luminous than the radio counterpart to \gw, and on very different timescales. Based on the radio monitoring by \citet{Kate}, the radio counterpart to \gw\ at the distance of even the most nearby known short GRBs would not have been detected.

\citet{Raff} and \citet{Kate} demonstrate that the temporal behavior of the X-ray and radio counterparts can be jointly explained by an off-axis jet. In particular, they find that for a jet opening angle of $15^{\circ}$, set to the median for short GRBs \citep{fbm+15}, the best-fit family of solutions has a range of circumburst densities, $\approx 10^{-4}-0.01$~cm$^{-3}$, jet energies of $10^{49}-10^{50}$~erg, and observer angles, $\theta_{\rm obs} \approx 20-40^{\circ}$, where the range of solutions is dominated by the uncertainty in the microphysical parameters \citep{Kate,Raff}. The range of allowed energies and densities inferred from the radio and X-ray counterparts to \gw\ is fully consistent with those inferred from on-axis short GRB afterglows, which have medians of $\approx 8 \times 10^{49}$~erg and $\approx (0.3-1.5) \times 10^{-2}$~cm$^{-3}$, respectively \citep{fbm+15}. This remarkable similarity hints that viewing angle effects are the dominant, and possibly the only, factor causing the difference in the observed radio and X-ray behavior between this event and on-axis short GRBs.

%%%%%%%%%% FIGURE
\begin{figure}
\tabcolsep0.0in
\includegraphics*[width=0.48\textwidth,clip=]{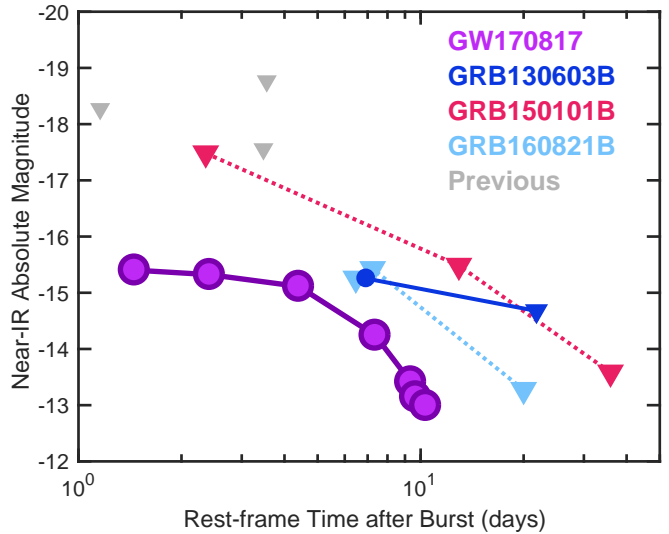}
\caption{A comparison of the luminosity and temporal evolution of the counterpart to \gw\ to previous claimed kilonovae and late-time searches following short GRBs. The rest-frame near-IR $H$-band light curves from Gemini-South observations of the counterpart to \gw\ are shown \citep{Phil}. Also shown is the near-IR excess following GRB\,130603B, interpreted as a kilonova (circles; \citealt{bfc13,tlf+13}), and $3\sigma$ upper limits following previous short GRBs (triangles; \citealt{fbm+15,fmc+16,kkl+17}, and this work). This comparison demonstrates that the near-IR excess following GRB\,130603B was $\approx 3$~times more luminous than the counterpart to \gw\ at a similar rest-frame time. This also demonstrates that near-IR searches following short GRBs were not sensitive enough or did not occur on the proper timescales to detect a kilonova with similar near-IR luminosity to that of \gw.
\label{fig:kn}}
\end{figure}
%%%%%%%%%% FIGURE

%%%%%%%%%%%%%%% TABLE
\begin{deluxetable*}{lcccc}[t]
\tablecolumns{5}
\tablewidth{0pc}
\tablecaption{Comparison of Stellar Population Properties
\label{tab:props}}
\tablehead{
\colhead {Property}	&
\colhead {SGRB SF median}	 &
\colhead {SGRB Ell. median} 		&
\colhead {SGRB total median} 		&
\colhead {NGC\,4993}		    \\
\colhead {(1)}	&
\colhead {(2)}	 &
\colhead {(3)} 		&
\colhead {(4)} 		&
\colhead {(5)}		    
}
\startdata
Stellar Mass (log($M_*/M_{\odot}$) & 9.65 (0.90) & 10.85 (0.33) & 10.10 (0.72) & 10.65 \\
Rest-frame $B$-band Luminosity ($L_B$/$L*$) & 0.70 (1.00) & 1.0 (0.80) & 0.85 (0.94) & 4.2 \\
Star Formation Rate ($M_{\odot}$~yr$^{-1}$) & 1.1 (0) & $\lesssim 0.2$ (0) & $\lesssim 0.5$ (0) & 0.01 \\
Stellar Population Age (Gyr) & 0.26 (1.00) & 1.3 (1.00) & 0.50 (1.00) & 11.2 \\
Projected Physical Offset $\delta R$ (kpc) & 5.5 (0.27) & 20.7 (0.17) & 6.9 (0.24) & 2.1 \\
Fractional Flux & 0.14 (0.75) & 0.21 (0.80) & 0.19 (0.76) & 0.54 \\
\enddata
\tablecomments{Columns correspond to: (1) Stellar population property, (2) Median for star-forming short GRB hosts, (3) Median for elliptical short GRB hosts, (4) Median for total short GRB host population, and (5) Value for NGC\,4993 from \citet{Peter}. In each column, numbers in parentheses represent the fraction of events in each class with values below that of NGC\,4993.}
\end{deluxetable*}
%%%%%%%%%%%%%%% TABLE

\subsection{Optical and Near-IR Kilonovae}

Next we compare the luminosity and temporal evolution of the optical/near-IR counterpart to \gw\ \citep{Marcelle,Phil,Chornock,Matt} to previous claims of kilonovae from short GRBs (Figure~\ref{fig:kn}). We also investigate whether previous searches following short GRBs were sensitive enough to detect excess emission with similar luminosities to the counterpart of \gw.

In the rest-frame optical band, the two claimed detections of kilonovae, GRB\,050709 \citep{jhl+16} and GRB\,060614 \citep{yjl+15,jlc+15}, are $\gtrsim 3-5$ times more luminous than the counterpart to \gw\ at $\gtrsim 3$~days. At near-IR wavelengths, the single detection following GRB\,130603B at $\approx 7$~days had a luminosity $\approx30$~times greater than the expected level of the afterglow at that time \citep{bfc13,tlf+13}. This significant near-IR excess, in conjunction with the extremely red color, represents the most convincing case of a kilonova following a cosmological short GRB. The comparison in Figure~\ref{fig:kn} demonstrates that the near-IR excess following GRB\,130603B was substantially more luminous, $\approx3$~times brighter, than the near-IR counterpart to \gw\ at the same epoch (Figure~\ref{fig:kn}). Moreover, all three claimed kilonovae remain luminous for longer after the merger in their respective rest-frame bands.

At the most basic level, the larger observed luminosities and sustained brightness for the three short GRBs could be explained by differences in the masses and velocities of ejected material during and subsequent to the mergers (e.g., \citealt{mmd+10,bk13}). For both GRB\,050709 and GRB\,060614, the inferred ejecta masses and velocities are $\approx 0.05-0.1\,M_{\odot}$ and $0.1-0.2c$ \citep{yjl+15,jhl+16}. Using single-component models available at the time \citep{bk13}, the luminosity of the kilonova following GRB\,130603B translated to an ejecta mass of $\approx0.03-0.08\,M_{\odot}$ for velocities of $0.1-0.3c$ \citep{bfc13}.

However, it has been suggested that there may be an additional early ``blue'' kilonova component as the result of lanthanide-free material from disk winds, amplified by the formation of hypermassive neutron star following the merger that is at least temporarily stable to collapse \citep{mf14,kfm15,met17}. Indeed, observations of the counterpart to \gw\ have demonstrated that single-component models are likely inadequate to explain kilonova behavior. The combination of multi-band optical/near-IR photometry of the counterpart to \gw\ \citep{Phil}, high-cadence UV/optical spectroscopy \citep{Matt}, and high-cadence near-IR spectroscopy \citep{Chornock}, show evidence for ``blue'' and ``red'' kilonova components, with overall lower inferred ejecta masses when compared to values inferred from short GRBs: $0.01\,M_{\odot}$ and $\approx 0.03-0.04\,M_{\odot}$, respectively. Still, the comparison between potential kilonovae following short GRBs and the optical/near-IR counterpart to \gw\ suggests that there may be a broad range of kilonova luminosities, colors, and timescales.

Finally, we utilize optical/near-IR upper limits following short GRBs to explore whether such searches would have been sensitive to the optical/near-IR counterpart to \gw, had it originated at higher redshifts. There are numerous optical limits following short GRBs at $\lesssim 1$~day from afterglow searches \citep{fbm+15}, and the deepest limit is following GRB\,050509B, with $M_r  \approx -14.3$ at $\approx\!0.9$ days \citep{csb+05,bpp+06}. This is $\approx 1$~mag deeper, or $\approx 2.2$~times less luminous, than the optical brightness of the counterpart to \gw\ \citep{Phil,Marcelle} interpolated to the same rest-frame time. Thus, a ``blue'' kilonova of similar luminosity to \gw\ can be confidently ruled out for GRB\,050509B (see also \citealt{hsg+05,mmd+10}), which also likely originated from an elliptical host galaxy at a low redshift of $z=0.225$ \citep{gso+05,bpp+06}. We note that since this event was detected as an on-axis short GRB, it is unlikely that there was significant obscuration of any ``blue'' component by a ``red'' kilonova component, which is expected to be more equatorial in geometry \citep{mf14}.

Optical and near-IR searches following all remaining short GRBs were not sensitive enough or did not occur on the proper timescales to detect a kilonova with similar luminosities to that of \gw. If the luminosity of the counterpart to \gw\ is representative of kilonovae from neutron star mergers, then searches following cosmological short GRBs will only probe the brightest end of the kilonova luminosity function. An added complication to additional kilonova detections is that $\approx 30\%$ of on-axis short GRBs have optical afterglow emission that dominates at early times, with a median of $M_r \approx -18.8$~mag at 10~hr \citep{ber14,fbm+15}. Thus, if the luminosity of \gw\ is representative of most kilonovae, the most promising route to accurately mapping the luminosity function is via Advanced LIGO/Virgo triggers.

%%%%%%%%%%%%%%% FIGURE
\begin{figure*}
\centering
\includegraphics*[width=0.45\textwidth,clip=]{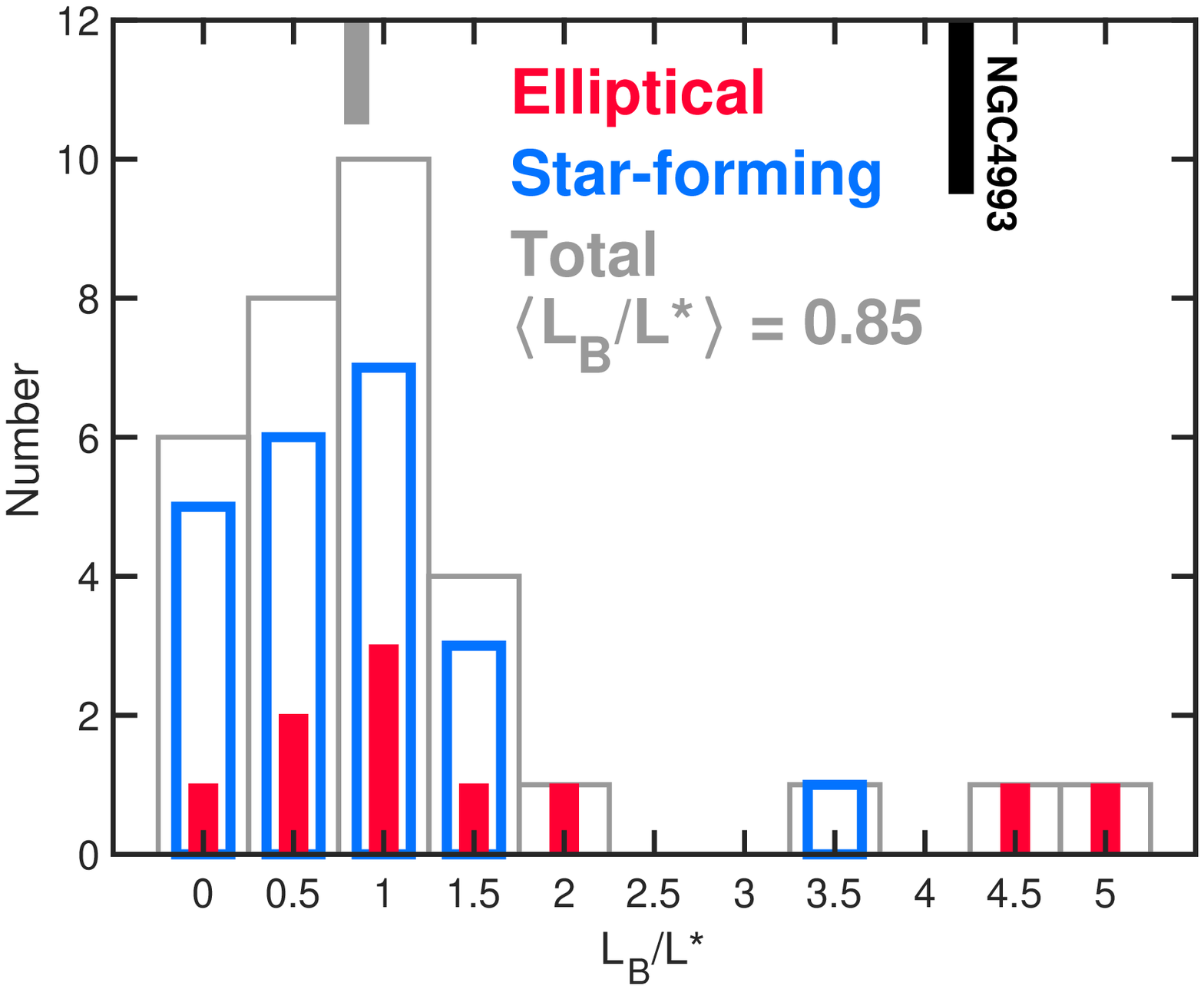} 
\includegraphics*[width=0.45\textwidth,clip=]{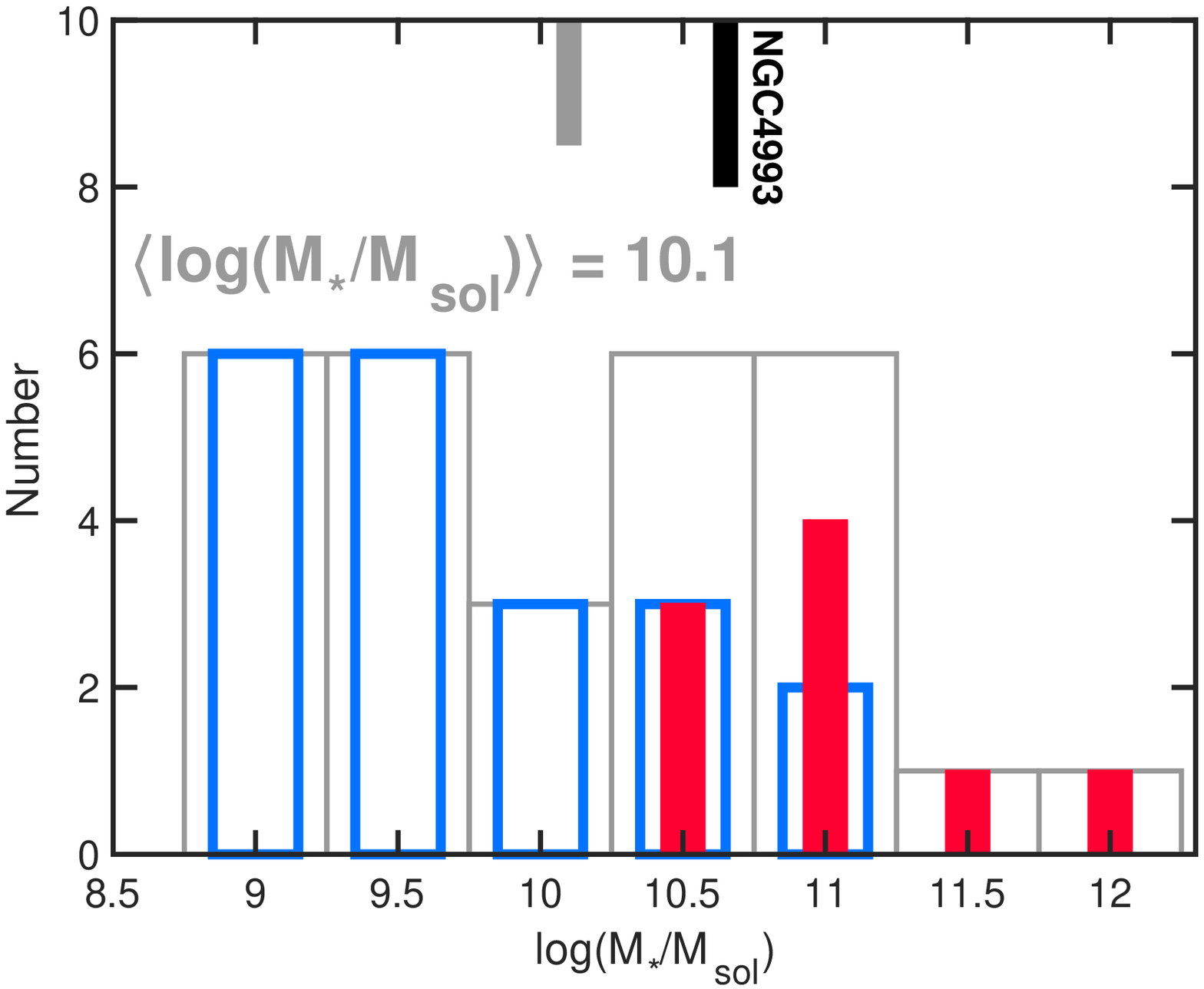} \\
\includegraphics*[width=0.45\textwidth,clip=]{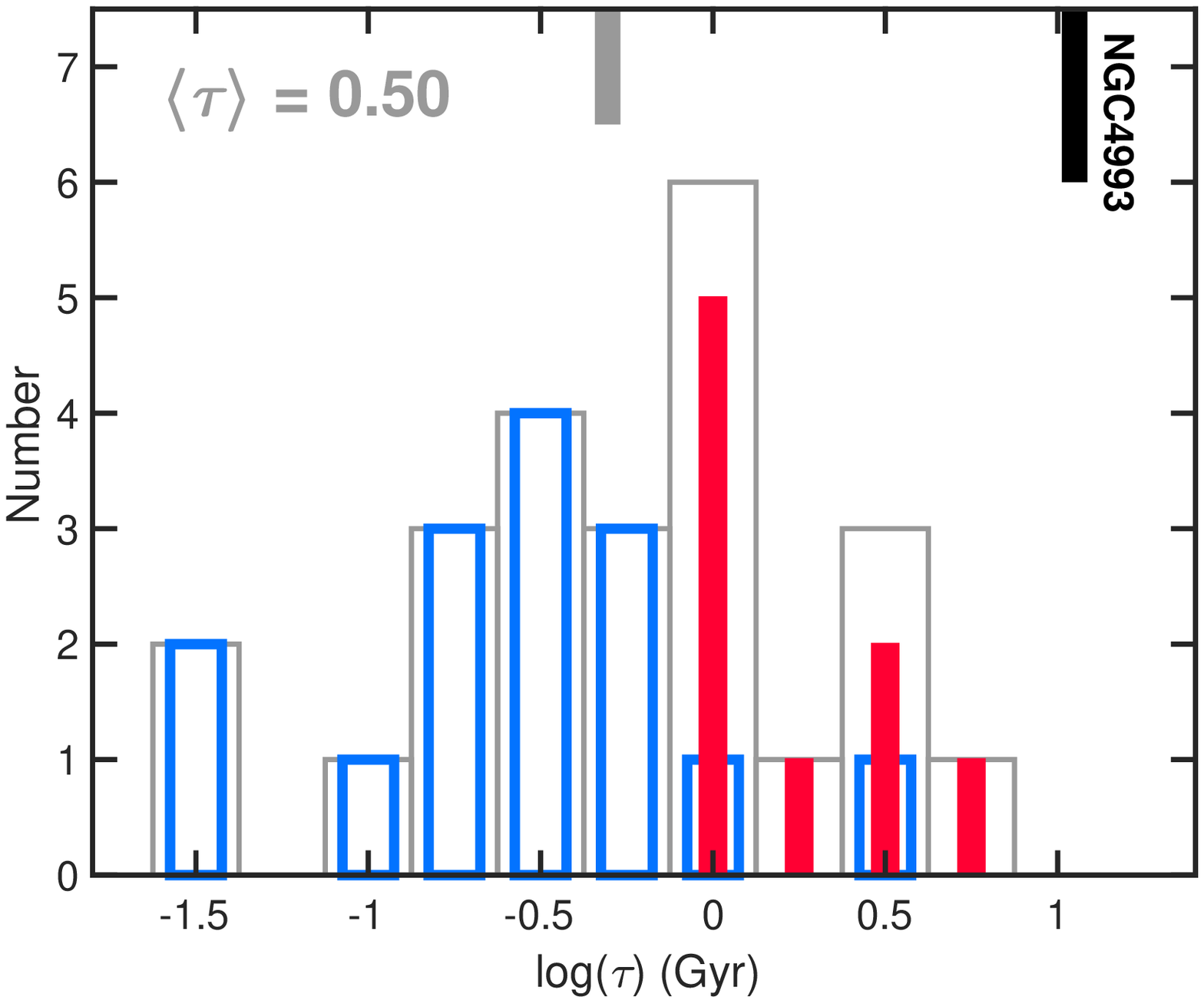} 
\includegraphics*[width=0.45\textwidth,clip=]{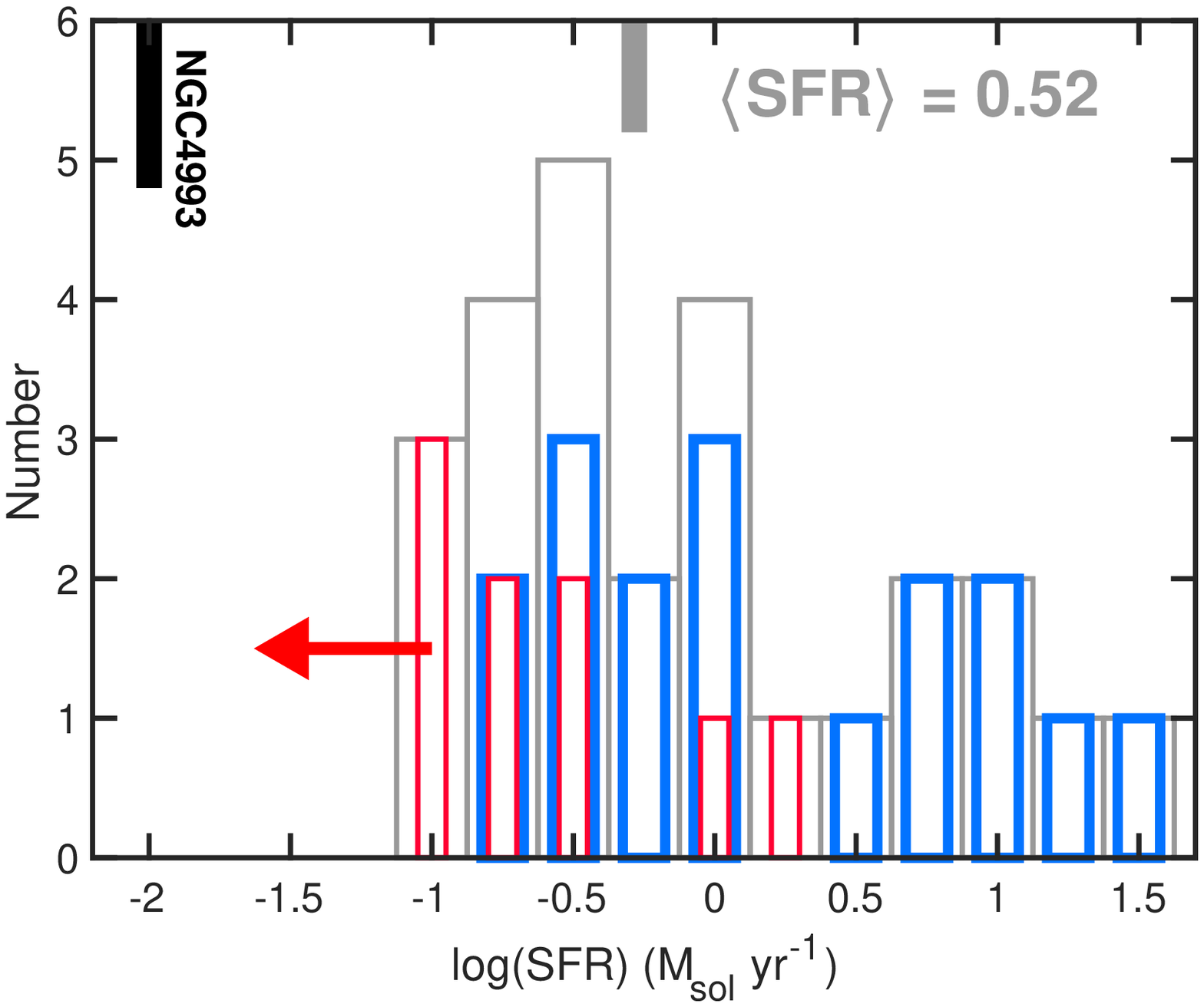} \\
\includegraphics*[width=0.45\textwidth,clip=]{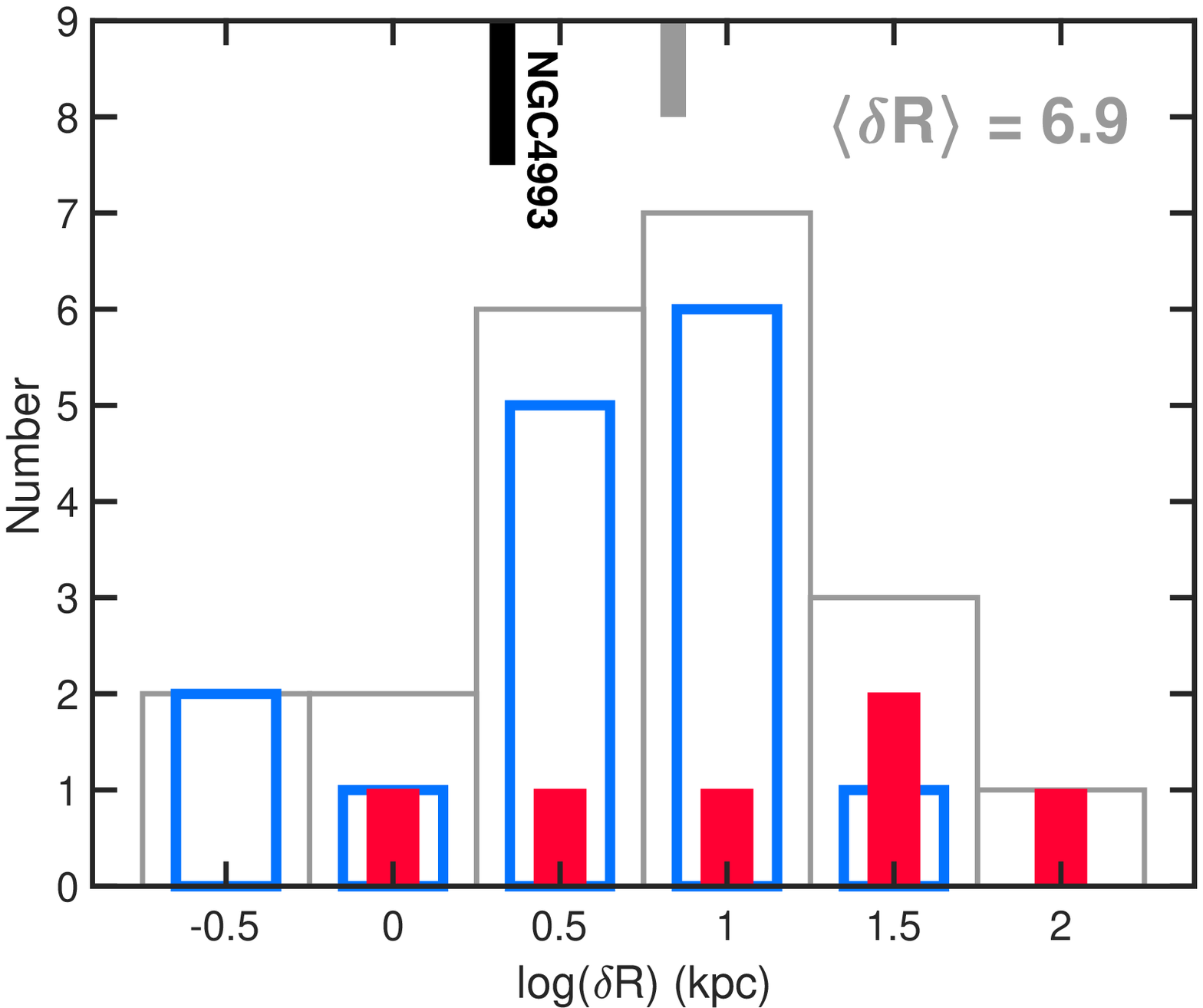} 
\includegraphics*[width=0.45\textwidth,clip=]{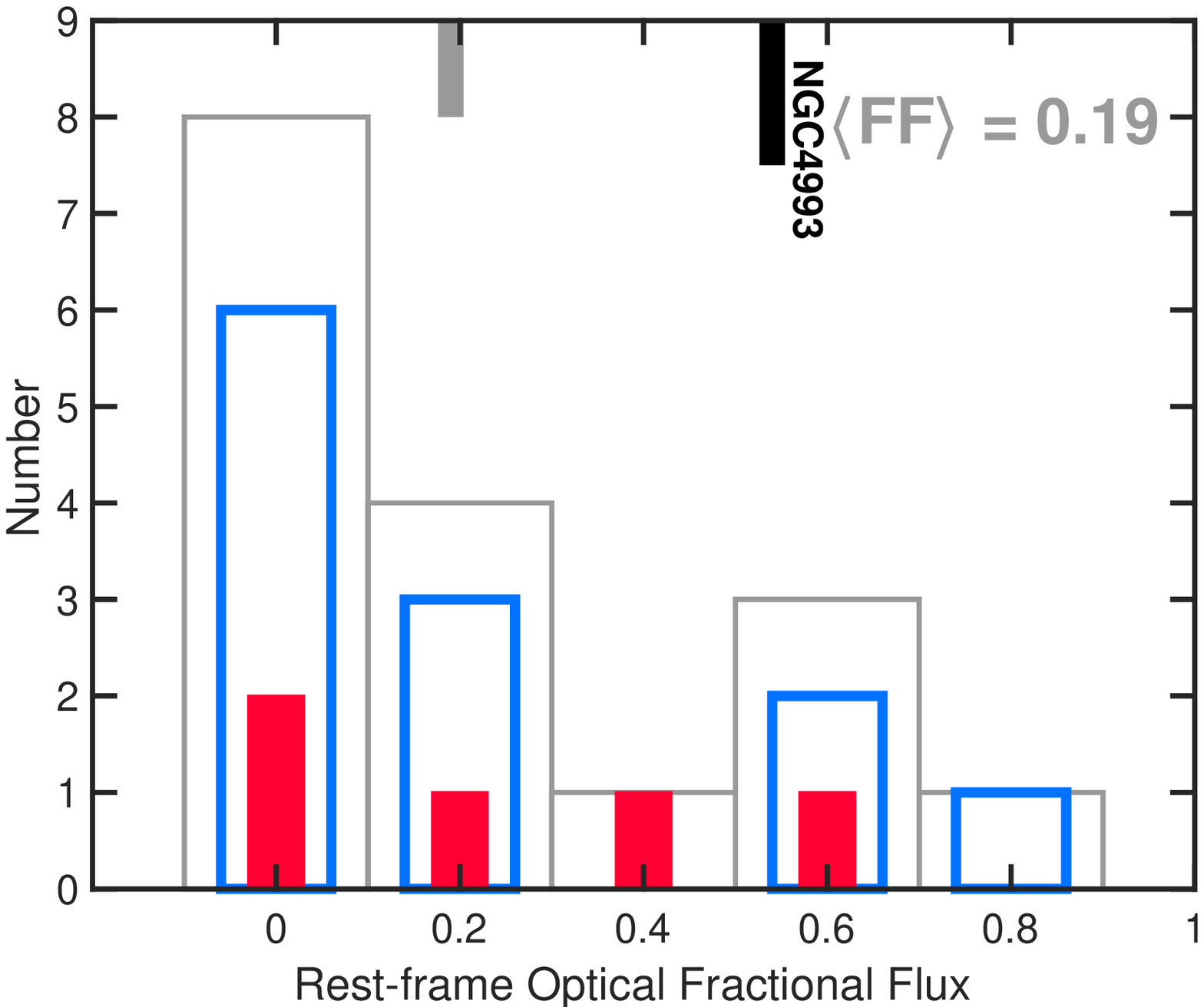} \\
\caption{Distributions of stellar population properties for the host galaxies of short GRBs discovered over 2004-2017: rest-frame $B$-band luminosity (top left), stellar mass (top right), stellar population age (middle left), star formation rate (middle right), projected physical offset (bottom left) and rest-frame optical fractional flux (bottom right). In each panel, the distributions for the total population (grey), star-forming hosts (blue) and elliptical hosts (red) are shown. The location of NGC\,4993 in each distribution (black line) along with the median for the short GRB host population (grey line) are marked. For the star formation rate panel, the open red histogram indicates upper limits measured on the star formation rates.
\label{fig:prophist}}
\end{figure*}
%%%%%%%%%%%%%%% FIGURE

\subsection{Location}

The locations of short GRBs with respect to their host galaxies are an important probe of their progenitor systems, as they contain key information about their local environments at the time of explosion. Past studies have shown that short GRBs have a median projected physical offset ($\delta R$) of $\approx 5$~kpc \citep{fbf10,fb13}, and span a wide range of $\approx 1-75$~kpc \citep{ber10,tlt+14}. Moreover, as a population, their locations are only very weakly correlated with their underlying host light distributions, indicating that they do not trace the distributions of their hosts' stellar mass or star formation. In general, their locations demonstrate that the progenitors must migrate from their birth sites to their explosion sites, and have been used as one of the primary arguments that they originate from NS-NS/NS-BH progenitors \citep{fbf10,ber10,fb13,tlt+14}.

In this vein, \citet{Peter} used {\it HST} optical and near-IR data and found that the counterpart to \gw\ is located at a projected physical offset of $\approx 2.1$~kpc from the center of \ngc. This value is low compared to short GRBs, with only $\approx 17-27\%$ of events located closer to their host centers, regardless of host galaxy type (Figure~\ref{fig:prophist} and Table~\ref{tab:props}). While the median offset for short GRBs in elliptical hosts is somewhat large, $\approx 21$~kpc, this is based on only six events which span the full range of $\approx 1-50$~kpc (Figure~\ref{fig:prophist}).

A complementary tool to physical offsets is to examine the location of the counterpart to \gw\ with respect to \ngc's light distribution (``fractional flux''). Using this method, \citet{Peter} found that the counterpart lies on a region of average brightness in the rest-frame optical band, with a fractional flux value of 0.54, indicating that there is nothing unusual about this position with respect to its host stellar mass distribution. However, this value is high compared to short GRBs, with $\approx 75-80\%$ of events on lower brightness levels (Figure~\ref{fig:prophist} and Table~\ref{tab:props}). Together with the small projected physical offset and old stellar population age of $\approx 11.2$~Gyr, this implies that the progenitor of \gw\ may have had a comparatively small kick velocity compared to those inferred for short GRBs, or happened to be on a part of its trajectory close to its host at the time of merger. However, since we have no {\it a priori} information on where the system was born, we can only place an upper limit on the kick velocity using the velocity dispersion of the host galaxy, which \citet{Peter} find to be $\lesssim 150-200$~km~s$^{-1}$. Furthermore, if the location of the counterpart to \gw\ is representative, NS-NS mergers may be more correlated to their host stellar mass distribution than implied from short GRBs alone.

\subsection{Stellar Population Properties}

Characterizing the host galaxy environments of short GRBs is key to understanding the parent stellar populations in which these explosive transients were born. From fitting the optical and near-IR SED of \ngc, \citet{Peter} found that \ngc\ has a stellar mass of log($M_*/M_{\odot})=10.65$, rest-frame $B$-band luminosity of $\approx 2 \times 10^{10}\,L_{\odot}$ ($4.2L^*$ when compared to the local galaxy luminosity function \citealt{bhb+03}), stellar population age of $\approx 11.2$~Gyr, and star formation rate of $0.01\,M_{\odot}$~yr$^{-1}$ (Table~\ref{tab:props}). Moreover, their analysis from surface brightness profile fitting indicates an elliptical morphology and half-light radius of $\approx 3-4$~kpc. Capitalizing on the past decade of short GRB host galaxy studies \citep{ber09,fbc+13,ber14}, we can now compare the stellar population of \ngc\ with the rest-frame $B$-band luminosities, stellar masses, stellar population ages and star formation rates of short GRB hosts.

The elliptical morphology along with the low level of ongoing star formation in \ngc\ is consistent with an early-type galaxy designation, which comprise only $\approx 1/3$ of short GRB hosts; the majority of short GRBs occur in star-forming galaxies \citep{ber09,fbc+13}. \ngc\ is among the most optically luminous host galaxies, super-ceded only by two other elliptical galaxies at low-redshift: GRB\,050509B ($z=0.225$; \citealt{gso+05,bpp+06}) and GRB\,150101B ($z=0.134$; \citealt{fmc+16}). However, its half-light radius is at the median level compared to short GRBs of $3.5$~kpc, suggesting a comparatively compact morphology.

\ngc\ also has by far the oldest stellar population age of $\approx 11.2$~Gyr, with the short GRBs in our sample spanning $\approx 0.03-4.4$~Gyr. Using stellar population age as a proxy for the delay time between the formation and merger of the progenitor system, this suggests a very broad range of delay times, as predicted from population synthesis models of NS-NS binaries \citep{bpb+06}. For a delay time distribution of the form $P(\tau) \propto \tau^n$, the stellar population ages of short GRBs, together with the relative burst rates in elliptical and star-forming galaxies, suggest that $n \approx -1$ \citep{zr07,obk08,lb10,fbc+13}. However, if there continues to be an increasing number of detected binary neutron star mergers with long delay times of $\gtrsim 5$~Gyr, the shape of the delay time distribution will shift toward values of $n \gtrsim -1$ (e.g., \citealt{zr07}). Additional events detected within the Advanced LIGO/Virgo volume will be necessary to provide a low-redshift anchor and delineate the delay time distribution.

In terms of star formation rates, short GRB star-forming hosts span a range of $\approx 0.1-30\,M_{\odot}$~yr$^{-1}$ \citep{ber09,pmm+11,bzl+13,ber14} with a median of $\approx 1.1\,M_{\odot}$~yr$^{-1}$, comparable to that of the Milky Way (Table~\ref{tab:props}). Short GRB hosts with no previous evidence of ongoing star formation had limits of $\lesssim 0.05-1.5\,M_{\odot}$~yr$^{-1}$ (\citealt{ber09,ber14}; Table~\ref{tab:props}). In this context, \ngc\ also has the lowest measured star formation rate of any short GRB host galaxy, enabled by its proximity, and is well below the most constraining limits for cosmological hosts. This low star formation rate is also commensurate with its old stellar population age \citep{Peter}.

Finally, the inferred stellar masses for the short GRBs in our sample have a median of log$(M_*/M_{\odot})\approx 10.1$ (Table~\ref{tab:props}). \ngc\ thus has a comparatively high stellar mass relative to the short GRB population, with log($M_*/M_{\odot})=10.65$, as $\approx 70\%$ of hosts have lower stellar masses. Compared to the stellar masses of elliptical galaxies, \ngc\ is well below average with only $\approx 1/3$ of short GRB elliptical hosts with lower stellar masses. 

In summary, \ngc\ is superlative in its large optical luminosity, old stellar population age, and low star-formation rate compared to the short GRB host population. It is average in terms of its size and above average in terms of its stellar mass. 

\section{Conclusions}
\label{sec:conc}

The past decade of short GRBs serve as a remarkable comparison sample to the counterpart and host galaxy of \gw. Here, we have provided a thorough comparison between the radio through X-ray counterpart to \gw\ and the afterglows and claimed kilonovae of short GRBs. Additonally, we have placed the host galaxy, \ngc, in the context of the galaxies hosting short GRBs. In particular, we conclude the following:

\begin{itemize}

\item Compared to on-axis short GRBs, the X-ray counterpart to \gw\ is $\approx 3000$ (50)~times less luminous than the median (faintest detected) X-ray afterglow luminosity. Similarly, the radio counterpart to \gw\ is $\gtrsim 10^{4}$ ($\gtrsim 500$)~times less luminous than the median (faintest detected) radio afterglows of short GRBs. Shifted to the distances of short GRBs, the counterpart to \gw\ would not have been detected given the capabilities of current X-ray and radio facilities.

\item Interpreted as an off-axis jet, the range of allowed energies and densities given by the radio and X-ray modeling \citep{Kate,Raff} is strikingly similar to those inferred for short GRBs. This suggests that viewing angle effects are the dominant, and perhaps only, difference between the observed radio and X-ray behavior of this event and on-axis cosmological short GRBs.

\item A comparison of the optical and near-IR counterpart to \gw\ to previous claimed kilonovae following short GRBs demonstrates that \gw\ is comparatively under-luminous by a factor of $\approx 3-5$, and becomes fainter more quickly, suggesting a range of kilonova luminosities and timescales. We further find that a ``blue'' kilonova of comparable optical luminosity to \gw\ can be ruled out in one previous event, GRB\,050509B. In addition, optical and near-IR searches following short GRBs were not sensitive enough or did not occur on the proper timescales to detect a kilonova with similar luminosities to that of \gw.

\item At a projected physical offset of $\approx 2.1$~kpc from the center of \ngc, the counterpart to \gw\ is relatively close to its host center, with only $\approx 25\%$~of short GRBs more proximal to their hosts. This is consistent with a relatively high fractional flux value, with $\approx 75\%$ of short GRBs on fainter rest-frame optical regions of their host galaxies. This implies that the progenitor of \gw\ may have had a comparatively small kick velocity compared to those inferred for short GRBs, or happened to be on a part of its trajectory close to its host at the time of merger. If the location of the counterpart to \gw\ is representative, NS-NS mergers may be more correlated to their host stellar mass distribution than implied from short GRBs alone.

\item In terms of host stellar population properties, \ngc\ is superlative in terms of its old stellar population age, low star formation rate, and large $B$-band luminosity when compared to the properties of short GRB hosts. \ngc\ is at the average level for its size and above average for its stellar mass. If the counterpart to \gw\ represents one of a missing population of observable NS-NS mergers with long inferred delay times, additional events within the Advanced LIGO/Virgo volume will be needed to provide a low-redshift anchor and quantify the delay time distribution.

\end{itemize}

For the first time, we are able to compare the cosmological population of short GRBs to a local analogue. An interesting possibility that arises from the detection of gamma-rays \citep{GBMdetection} for an local, off-axis event (as inferred from radio and X-rays; \citealt{Kate,Raff}) is that previously-detected, weak short GRBs are in fact local, off-axis events. Indeed, \citet{tcl+05} examined a population of BATSE short GRBs that were not well localized and thus did not have redshift determinations, to explore any correlations with local galaxies. They found that up to $\approx 1/4$ of short GRBs may originate at local distances of $\lesssim 100$~Mpc. The fraction of previously-detected weak short GRBs that are actually local, off-axis events will be solidified with additional joint detections of NS-NS/NS-BH mergers from Advanced LIGO/Virgo and gamma-ray transients.

Over the next several years, using the incoming flow of local events from Advanced LIGO/Virgo, we will be able to infer the properties of their jets, delineate the kilonova luminosity distribution, determine the rates of binary neutron star mergers, and infer the properties of their local host galaxies. Such studies will help to define the key similarities, and differences, between the local population of compact object mergers and their cosmological counterparts.

\section*{Acknowledgments}
\noindent WF acknowledges support for Program number HST-HF2-51390.001-A, provided by NASA through a grant from the Space Telescope Science Institute, which is operated by the Association of Universities for Research in Astronomy, Incorporated, under NASA contract NAS5-26555. The Berger Time-Domain Group at Harvard is supported in part by the NSF through grants AST-1411763 and AST-1714498, and by NASA through grants NNX15AE50G and NNX16AC22G. DAB is supported by NSF award PHY-1707954. Based on observations made with the NASA/ESA Hubble Space Telescope, obtained from the data archive at the Space Telescope Science Institute. STScI is operated by the Association of Universities for Research in Astronomy, Inc. under NASA contract NAS 5-26555. This work made use of data supplied by the UK Swift Science Data Centre at the University of Leicester. The National Radio Astronomy Observatory is a facility of the National Science Foundation operated under cooperative agreement by Associated Universities, Inc.

%\bibliographystyle{apj}
%\bibliography{journals_apj,refs}

\end{document}